\RequirePackage[2024-02-02]{latexrelease}
\documentclass[aps,prc,twocolumn,floatfix,superscriptaddress]{revtex4}
\usepackage{epsfig}
\usepackage{amsmath}
\usepackage{amssymb} 
\usepackage{url}
\usepackage{color}
\usepackage{graphicx} 

\newcommand{\pT}{p_{\mathrm{T}}}
\newcommand{\vzpT}{v_0(\pT)}
\newcommand{\snn}{\sqrt{s_{\mathrm{NN}}}}
\newcommand{\nch}{N_{\mathrm{ch}}}
\newcommand{\ptref}{p_{\mathrm{T}}^{\mathrm{ref}}}
\newcommand{\etagap}{\eta_{\mathrm{gap}}}
\newcommand{\mT}{m_{\mathrm{T}}}
\newcommand{\betaT}{\langle\beta_{\mathrm{T}}\rangle}

\begin{document}

\title{Characterizing $\pT$-differential Radial-Flow Fluctuations in Au+Au Collisions at FAIR with a Multi-Phase Transport Model}

\author{Sanchari Thakur}
\email{sanchari.thakur@vecc.gov.in}

\affiliation{Variable Energy Cyclotron Centre, 1/AF, Bidhan Nagar, Kolkata 700064, India}

\author{Partha Pratim Bhaduri}
\email{partha.bhaduri@vecc.gov.in}

\affiliation{Variable Energy Cyclotron Centre, 1/AF, Bidhan Nagar, Kolkata 700064, India}
\affiliation{Homi Bhabha National Institute, Training School Complex, Anushaktinagar, Mumbai 400094, India}

\begin{abstract}
We report the first predictions of the transverse-momentum ($\pT$)-differential radial-flow fluctuation observable, $\vzpT$, in Au+Au collisions at the top SIS100 energy, $\snn=4.9$~GeV, which will be available at the FAIR accelerator complex. Radial flow and its fluctuations are studied for different centrality intervals and reference-$\pT$ selections to investigate the properties and robustness of the calculated long-range correlation. $\vzpT$ displays a characteristic evolution from negative values at low $\pT$ to positive values at higher $\pT$. The integrated $v_0$, obtained with different reference-$\pT$ intervals, follows a common trend, with an approximately inverse-square-root dependence on the inclusive charged-particle multiplicity ($\nch$). The $\vzpT$ signal changes little when the pseudorapidity gap is varied between 0.2 and 0.8, or when the edges of the reference-$\pT$ interval are varied, in the bulk region. For identified particles, the pion and proton $\vzpT$ are ordered at low $\pT$, qualitatively similar to the mass ordering seen at higher energies. However, unlike at higher collision energies, $\vzpT$ does not exhibit constituent-quark number scaling, so the results do not support an interpretation in terms of partonic collectivity. The results are also in partial contradiction with the behavior of hadronic elliptic flow in these low-energy collisions. These results provide constraints on the evolution of radial correlations toward lower collision energies and motivate comparisons with complementary observables to clarify their connection to collective dynamics.
\end{abstract}


\maketitle

\section{Introduction}

Relativistic heavy-ion collisions at (top) RHIC and the LHC energies produce a deconfined quark--gluon plasma (QGP) at high temperature ($T$) and nearly vanishing baryon chemical potential ($\mu_B$)~\cite{qgp1,qgp2,qgp3,qgp4,qgp5}. Lowering the collision energy drives the produced fireball toward large $\mu_B$, the region of the QCD phase diagram targeted by present and planned facilities worldwide~\cite{Galatyuk:2019lcf}. One of these is the Compressed Baryonic Matter (CBM) experiment, a fixed-target experiment under construction at the Facility for Antiproton and Ion Research (FAIR), which will study heavy-ion collisions in the energy range $\snn = 2.7$--4.9~GeV to map strongly interacting matter at high net-baryon density and moderate temperature~\cite{CBM2017,CBM2026}. Collective flow is a central part of the CBM physics program because it constrains the equation of state (EoS) and the transport properties of dense QCD matter, and the characteristics of anisotropic flow expected at FAIR energies have already been studied in several models~\cite{Bhaduri:2010wi,Sarkar:2017fuy,Sarkar:2017fni,Rode:2019pey,Kundu:2021afz,Kundu:2022tec}.

Collective behavior shows up as long-range correlations among the produced particles in azimuth ($\phi$) and pseudorapidity ($\eta$)~\cite{coll1,coll2,coll3,coll4,coll5}. In the expanding medium, pressure gradients turn the spatial anisotropy of the initial state into an anisotropy in momentum space~\cite{coll1,coll2,coll6}, which has been measured for many collision systems and energies at RHIC and the LHC~\cite{flow1,flow2,flow3,flow4,v03,flow5}. It is quantified by the Fourier coefficients $v_n$ of the azimuthal distribution with respect to the reaction plane, which also receive nonflow contributions from jets, resonance decays, and other few-particle correlations. The azimuthally symmetric component of the expansion, radial flow~\cite{radial}, pushes particles to higher transverse momentum ($\pT$), so that the measured hadron spectra reflect both the thermal motion of the source and its collective transverse velocity. Radial flow is traditionally extracted from Boltzmann-Gibbs blast-wave (BGBW) fits to identified-hadron spectra~\cite{blast3,blast1,blast2}. Such fits yield a single $\pT$-integrated mean transverse velocity, $\betaT$, and the inclusive spectra they analyze also include contributions from nonflow correlations.

A $\pT$-differential, nonflow-suppressed handle on radial flow is provided by $\vzpT$, recently introduced in Ref.~\cite{v01} and systematically developed in Ref.~\cite{v02}. It correlates, event by event, the fraction of particles in a given $\pT$ bin with the mean transverse momentum measured in a second subevent separated by a pseudorapidity gap $\Delta\eta$. The gap removes most short-range correlations, so that radial flow can be analyzed in the same manner as anisotropic flow. State-of-the-art hydrodynamic calculations~\cite{v01,v02,hydro} predict that $\vzpT$ is negative at low $\pT$ and positive at high $\pT$, since an event with a larger mean $\pT$ has a harder spectrum, and that identified hadrons show a mass ordering similar to that of $v_2(\pT)$. Because $\vzpT$ follows the isotropic expansion, it is expected to be sensitive mainly to the bulk viscosity ($\zeta/s$) and the equation of state (EoS) driving the expansion dynamics, and only weakly to the shear viscosity ($\eta/s$).

Both ALICE~\cite{radial} and ATLAS~\cite{v03} have measured $\vzpT$ in Pb+Pb collisions at $\snn = 5.02$~TeV. The ATLAS charged-particle data change sign near $\pT \approx 1$~GeV/$c$ and, for $\pT \lesssim 3$~GeV/$c$, are nearly independent of the pseudorapidity gap and of the reference-$\pT$ range and have a centrality-independent shape, which together point to a collective origin. ALICE observed a mass ordering of pions, kaons, and protons below $\pT \approx 3$~GeV/$c$ and larger proton values above it, and found approximate number-of-constituent-quark (NCQ) scaling~\cite{scaling1,scaling2,scaling3,scaling4,scaling5} when $\vzpT/n_q$ is plotted against $(\mT-m_0)/n_q$, suggesting that the collectivity develops at the partonic stage. Hydrodynamic calculations reproduce the ALICE data up to $\pT \approx 2$~GeV/$c$ and show that $\vzpT$ is sensitive to $\zeta/s$, the underlying QCD EoS, and initial-state fluctuations, but hardly to $\eta/s$~\cite{radial}.

Beyond the LHC, preliminary STAR results cover Au+Au collisions at $\snn = 7.7$--200~GeV~\cite{STARprelim}, and model studies have addressed the evolution of $\vzpT$ at RHIC energies~\cite{hydro,radial2,DuPRL}. In particular, Ref.~\cite{radial2} used the AMPT model at $\snn = 200$~GeV and found the mass ordering and baryon--meson splitting seen at the LHC, together with NCQ scaling in central collisions. On the other hand, in Ref.~\cite{hydro}, the author has studied the collision energy dependence of the characteristics of $\vzpT$ in $\snn = 0.2$--5.02~TeV heavy-ion collisions, using a multistage hybrid hydrodynamic model, implemented within the JETSCAPE framework~\cite{Putschke:2019yrg}. The response of $\vzpT$ to a wide range of physical effects and modeling configurations has been investigated across three collision energies. With decreasing beam energy, a more complex interplay of shear and bulk viscous forces was found to determine the 
shape of $\vzpT$. In Ref.~\cite{Parida:2026kyo}, the author has studied the influence of finite baryon density and baryon diffusion on event-wise mean $p_{T}$ fluctuations in $\snn=19.6$ GeV Au+Au collisions. To date, no study of $\vzpT$ has been carried out at FAIR energies, $\snn \lesssim 5$~GeV, where the produced medium is highly baryon-rich, and it is not known whether these features persist there. In this work, we present the first predictions of $\vzpT$ for Au+Au collisions at $\snn = 4.9$~GeV, the top SIS100 energy, using the string-melting version of the AMPT model (AMPT-SM), which over the last two decades has emerged as a useful tool to reproduce a variety of experimental observables in relativistic heavy-ion collisions in the high-baryon-density regime~\cite{Jena,Li:2025lix,Gao:2025voq,Zhu:2025kud,Li:2026kyw}. 
We follow the analysis procedure of Refs.~\cite{v02,radial2}, so that the results can be compared directly with those at higher energies.

The paper is organized as follows. For completeness, Section~II briefly describes the main features of the AMPT model, and Section~III presents the observables used in this work. Section~IV presents and discusses the results, followed by a summary of the main findings in Section~V.

\begin{figure*}[t]
\centering
\includegraphics[width=\textwidth]{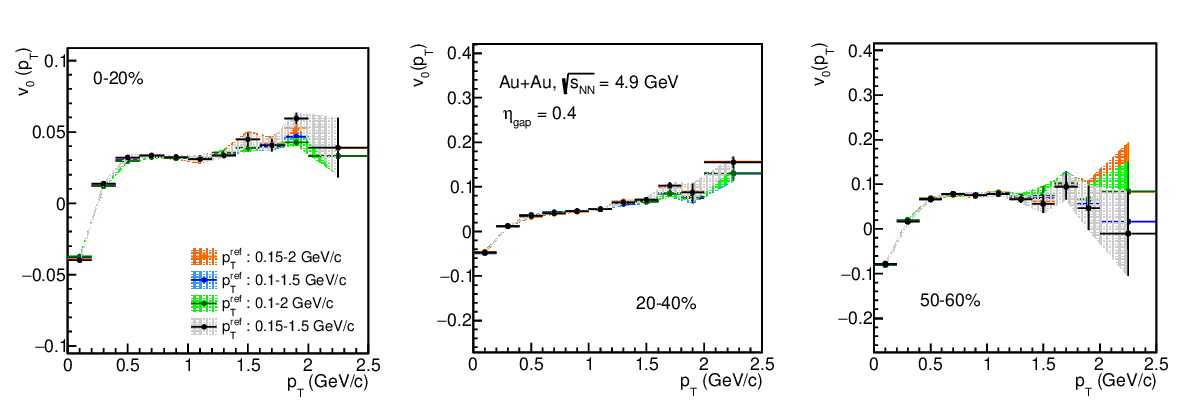}
\caption{(Color online) Transverse-momentum dependence of the radial-flow fluctuation observable $\vzpT$ in Au+Au collisions at $\snn=4.9$~GeV for the 0--20\%, 20--40\%, and 50--60\% centrality classes, for four reference-$\pT$ intervals and $\etagap = 0.4$.}
\label{eps1}
\end{figure*}

\section{AMPT model}

A Multi-Phase Transport (AMPT) model~\cite{ampt} is a Monte Carlo (MC) transport event generator extensively employed to investigate relativistic heavy-ion collisions at both RHIC and LHC energies. The model incorporates several stages of the collision dynamics, including the initial-state configuration, partonic interactions, hadron formation, and subsequent hadronic rescattering. Initial conditions are generated using the Heavy Ion Jet Interaction Generator (HIJING)~\cite{hij1,hij2}, which provides the spatial and momentum distributions of minijet partons as well as excited soft strings. The partonic stage is described by Zhang's Parton Cascade (ZPC)~\cite{zpc}, which treats interactions among quarks and gluons through elastic two-body scattering with parametrized cross sections.

Following the partonic evolution, hadronization is implemented according to the selected AMPT configuration. In the string-melting (SM) mode, nearby partons are combined through a quark-coalescence mechanism to form mesons and baryons. This approach is particularly useful for modeling a medium dominated by deconfined partonic degrees of freedom and is therefore commonly employed in studies of QGP properties. In the default AMPT configuration, hadronization instead proceeds through Lund string fragmentation. After hadron formation, the resulting particles undergo further interactions and rescattering, which are simulated using the relativistic transport model ART~\cite{art}. Through this multi-stage treatment, AMPT provides a framework for studying the collective behavior and dynamical evolution of matter produced in high-energy nuclear collisions.

In this work, we use the AMPT-SM version with the Lund string fragmentation function $f(z) \propto z^{-1}(1-z)^{a}\exp(-b\,\mT^{2}/z)$, with parameters $a = 0.55$ and $b = 0.15~\mathrm{GeV}^{-2}$. Here $z$ denotes the light-cone momentum fraction of the produced hadron of transverse mass $\mT$ relative to that of the fragmenting string. The parton--parton scattering cross section is modeled as $\sigma_{\rm parton} = 9\pi\alpha_{s}^{2}/(2\mu^{2})$, where $\alpha_{s}$ denotes the QCD coupling constant and $\mu$ the parton screening mass. In the present work, we use $\alpha_{s} = 0.33$ and $\mu = 2.265~\mathrm{fm}^{-1}$, corresponding to a parton scattering cross section of 3~mb. This set of parameters has been successful in reproducing particle production at RHIC energies down to $\snn = 7.7$~GeV~\cite{Jena}.
\begin{figure*}[t]
\centering
\includegraphics[width=\textwidth]{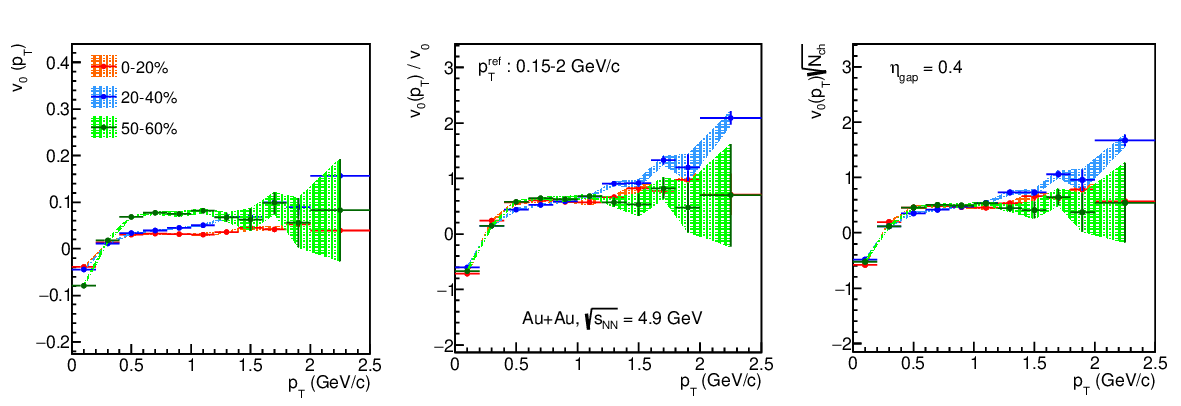}
\caption{(Color online) Transverse-momentum dependence of the radial-flow fluctuation observable $\vzpT$ for inclusive charged particles in Au+Au collisions at $\snn=4.9$~GeV. Results are shown for the 0--20\%, 20--40\%, and 50--60\% centrality classes. The left, middle, and right panels show $\vzpT$, $\vzpT/v_0$, and $\vzpT\sqrt{\nch}$, respectively.}
\label{eps2}
\end{figure*}

\section{Observables}

Following Refs.~\cite{v01,v02,radial2}, $\vzpT$ is obtained by correlating the $\pT$ spectrum measured in one pseudorapidity window with the mean transverse momentum measured in another. Final-state particles are binned in $\pT$, and the normalized covariance is taken between the fraction of particles in each bin and the event-wise mean $\pT$:

\begin{equation}
    v_0 (\pT) = \frac{\langle F_A^{\pT} [\pT]_B \rangle - \langle F_A^{\pT} \rangle \langle [\pT]_B \rangle}{\langle F_A^{\pT} \rangle \sigma_{[\pT]}}.
\label{Eq1}
\end{equation}

Here $A$ and $B$ are two pseudorapidity windows separated by a gap $\Delta\eta$ (denoted $\etagap$ in the figures), which suppresses short-range correlations between them. For each event, $F_A^{\pT}$ is the fraction of the charged particles in window $A$ that lie in a given $\pT$ bin,
\begin{equation}
  F_A^{\pT} = \frac{N_{\mathrm{ch}}^A (\pT)}{N_{\mathrm{ch}}^A},
\end{equation}
with $N_{\mathrm{ch}}^A (\pT)$ the number of charged hadrons in that bin and $N_{\mathrm{ch}}^A$ the total charged-particle multiplicity in $A$. The quantity $[\pT]_B$ in Eq.~(\ref{Eq1}) is the mean transverse momentum of the $N_{\mathrm{ch}}^B$ charged particles in window $B$ of the same event,
\begin{equation}
    [\pT]_B = \frac{1}{N_{\mathrm{ch}}^B}\sum_{i\in B} p_{\mathrm{T},i}.
\end{equation}
Angular brackets $\langle\ldots\rangle$ denote averages over the events of a centrality class, and $\sigma_{[\pT]}$, the standard deviation of the event-wise mean $\pT$, is estimated from the covariance between the two windows,
\begin{equation}
    \sigma_{[\pT]} = \sqrt{\langle [\pT]_A [\pT]_B \rangle - \langle[\pT]_A \rangle \langle[\pT]_B \rangle}.
\end{equation}

The $\pT$-integrated counterpart, $v_0$, measures the relative event-by-event fluctuation of the mean transverse momentum~\cite{v03,v01}:
\begin{equation}
    v_0 = \frac{\sqrt{\langle (\delta[\pT])^2\rangle_R}}{\langle [\pT] \rangle_R},
\end{equation}
where $\delta[\pT] = [\pT] - \langle [\pT] \rangle_R$ and $\langle\ldots\rangle_R$ denotes an average over events, with particles restricted to the reference transverse-momentum interval $R$.

The physical picture behind $\vzpT$ is that radial-flow fluctuations change the slope of the $\pT$ spectrum from event to event. In hydrodynamics, the same pressure gradients generate both the azimuthal and the long-range $\pT$ correlations: the initial geometry sets the anisotropic flow, whereas the initial size and temperature of the fireball set the spectral shape. Bulk viscosity opposes isotropic expansion, so $\vzpT$ is expected to be sensitive to it. The relation between $\vzpT$ and the blast-wave parameter $\betaT$~\cite{blast3} has been worked out in Ref.~\cite{blast2}: letting $\beta_{\mathrm{T}}$ fluctuate from event to event with a width $\sigma(\beta_{\mathrm{T}})$ produces a nonzero $\vzpT$ whose magnitude depends on kinetic freeze out parameters ($\beta_{\mathrm{T}}$ and $T_{kin}$) and their fluctuations, and $\vzpT$ vanishes if $\beta_{\mathrm{T}}$ does not fluctuate. 

The boost-invariant version of the blast-wave model, as employed in Ref.~\cite{blast2}, is not suitable in the FAIR energy domain because of the explicit breaking of longitudinal boost invariance in low-energy collisions. A possible solution is to employ a non-boost-invariant blast-wave model~\cite{Rode:2018hlj,Rode:2020vhu,Rode:2023vjl} to study the connection between the $\pT$-integrated and $\pT$-differential radial-flow parameters. We leave such a study to future work.

\section{Results}

\subsection{Transverse-momentum dependence of $\vzpT$}

We now present the results obtained from the AMPT model.
We begin with the $\pT$ dependence of $\vzpT$ of charged hadrons, in Au+Au collisions at $\snn=4.9$~GeV for three centrality intervals, 0--20\%, 20--40\%, and 50--60\%, shown in Fig.~\ref{eps1}. The observable is evaluated within $|\eta|\leq1$ using a pseudorapidity gap of $\etagap=0.4$, the standard value used by ALICE~\cite{radial}. The evaluation of $v_{0}(p_{T})$ is sensitive to the implemented kinematic cuts in the experimental analysis~\cite{DuPRL,Parida:2026uak}. In the present theoretical simulations, we have considered particles down to $p_{T} \rightarrow 0$. Correction for experimental bias introduced by detector-specific kinematic cuts will be necessary as and when measurements from FAIR become available for a meaningful model-to-data comparison~\cite{Parida:2026uak}. For our present study, four reference-$\pT$ intervals are considered: $0.15<\ptref<2.0$, $0.15<\ptref<1.5$, $0.10<\ptref<2.0$, and $0.10<\ptref<1.5$~GeV/$c$. This choice allows us to examine the sensitivity of the differential observable to the reference momentum interval. Defining centrality classes by impact parameter differs from the experimental procedure, in which centrality classes are determined a posteriori from the charged-particle multiplicity distribution. In the present exploratory study, to reduce the computational cost, we adopt the impact-parameter range corresponding to each centrality class from a Glauber model analysis~\cite{impact1}.

A characteristic $\pT$ dependence of $\vzpT$ is observed in all three centrality classes, similar to that found at higher collision energies in both simulations and measurements~\cite{radial,radial2}. At very low $\pT$, $\vzpT$ is negative; it crosses zero at around $\pT \simeq 0.3$~GeV/$c$ and rises at intermediate to high $\pT$ in all centrality classes. The negative values at low $\pT$ can be understood from the anticorrelation between event-by-event fluctuations of the mean transverse momentum and changes in the shape of the single-particle $\pT$ spectrum~\cite{v01,v02}. Events with a larger-than-average $[\pT]$, corresponding to stronger transverse expansion, tend to have a harder spectrum. Consequently, the relative particle yield at low $\pT$ is reduced, producing a negative contribution to $\vzpT$. At higher $\pT$, the same spectral hardening results in an enhanced yield and hence a positive $\vzpT$. This behavior is expected for a radial-flow fluctuation observable from simulations and measurements at higher energies. The different reference-$\pT$ selections give very similar trends over the low- and intermediate-$\pT$ regions. This agreement indicates that the dominant correlation in this momentum range is relatively insensitive to the precise choice of reference window.

The shape found at FAIR energy resembles that of the charged-hadron $\vzpT$ measured by ALICE in Pb+Pb collisions at $\snn=5.02$~TeV~\cite{radial}, but on a compressed $\pT$ scale. In our calculation, $\vzpT$ grows almost linearly with $\pT$ up to about 0.5~GeV/$c$ in all centrality classes, whereas the ALICE data stay linear up to about 2~GeV/$c$, even in peripheral collisions, in agreement with hydrodynamic calculations. At higher $\pT$, the LHC data fall below the hydrodynamic curves, as anisotropic flow does; ALICE interprets this as the onset of a regime in which hard scatterings and jet--medium interactions modify the spectrum and its fluctuations~\cite{radial}. ATLAS observed the same behavior and, by comparing four analysis procedures, showed that fluctuations of the total multiplicity shift $\vzpT$ vertically and move its zero crossing without changing its shape~\cite{v03}.

At the LHC, HIJING, which has no collective expansion but includes minijets, resonance decays, and nuclear effects, describes $\vzpT$ in peripheral collisions up to high $\pT$, reflecting the growing weight of hard processes there~\cite{radial}. In AMPT-SM, HIJING provides the initial conditions; no hydrodynamic predictions of $\vzpT$ exist yet at FAIR energies. In our results, the departure from linearity sets in at a much lower $\pT$ than at the LHC, as expected for a center-of-mass energy roughly three orders of magnitude lower. Likewise, the zero crossing moves from $\pT \sim 1$~GeV/$c$ at the LHC~\cite{v03} to $\pT \simeq 0.3$~GeV/$c$ here.

\begin{figure*}[t]
\centering
\includegraphics[width=0.8\textwidth]{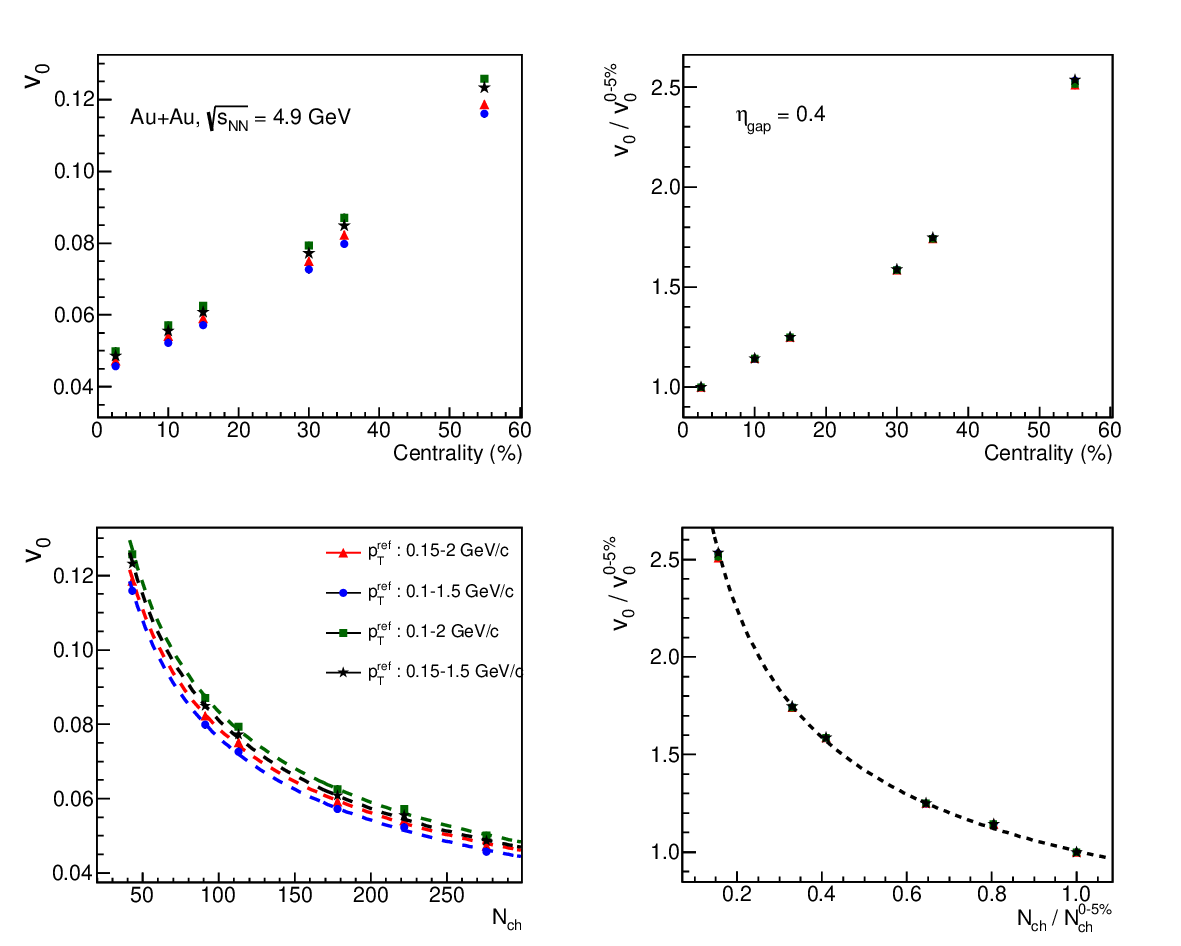}
\caption{(Color online) Centrality and charged-particle multiplicity dependence of the integrated radial-flow fluctuation observable $v_0$ in Au+Au collisions at $\snn=4.9$~GeV.}
\label{eps3}
\end{figure*}

\subsection{Centrality dependence of $\vzpT$}
Having discussed the $\pT$ dependence of $\vzpT$, we now investigate its dependence on collision centrality. Fig.~\ref{eps2} (left) shows $\vzpT$ for inclusive charged particles in three centrality intervals, 0--20\%, 20--40\%, and 50--60\%, evaluated for the reference-$\pT$ interval 0.15--2.0~GeV/$c$. The 0--20\% class shows a relatively smooth and modest $\pT$ dependence compared with the other centrality classes, which have larger fluctuations toward higher $\pT$. This is qualitatively expected, since the smaller particle multiplicity in peripheral events provides less statistical averaging of the event-wise $p_{T}$ fluctuations, thereby increasing the magnitude of the relative fluctuations.
Similarly, although central collisions produce larger, more extended systems with stronger average radial expansion, the relative size of event-by-event fluctuations is reduced because of the larger number of participating particles. The larger multiplicity therefore leads to stronger statistical averaging of fluctuations, resulting in a smaller magnitude of $\vzpT$. The enhanced fluctuations at larger $\pT$ indicate a region where non-collective contributions can become increasingly important.
Fig.~\ref{eps2} (middle) presents the corresponding scaled observable, $\vzpT/v_0$, for the different centrality classes, which is predicted to be insensitive to initial-state effects~\cite{v01}. Here $v_0$ represents the magnitude of the $\pT$-integrated radial flow.
This scaling reduces the overall centrality dependence of the radial-flow fluctuations up to about 1~GeV/$c$ and allows the shapes of the $\pT$ dependence to be compared directly between centralities. Qualitatively, the observed patterns are similar to those obtained in Pb+Pb and Au+Au studies at higher energies~\cite{v03,radial2}. At $\snn = 4.9$~GeV, $\vzpT/v_0$ is approximately independent of centrality at low $\pT$, as expected for a common response to radial-flow fluctuations, whose shape is predicted to be nearly independent of system size at a given collision energy~\cite{v02}. The observed scaling behavior is similar to that seen for anisotropic flow, $v_n(\pT)/v_n$, by the ATLAS Collaboration at the LHC~\cite{ATLAS:2018ezv}. However, as mentioned above, $\vzpT/v_0$ is mostly sensitive to bulk viscosity, whereas $v_n(\pT)/v_n$ depends on both the shear and bulk viscosities of the produced medium~\cite{radial,hydro}.
The progressively larger deviations from scaling at high $\pT$, especially in peripheral collisions, indicate imperfect scaling due to the increasing impact of non-collective contributions. Note that a minor deviation from hydrodynamic scaling is also observed at the LHC, particularly in peripheral collisions for $\pT > 5$~GeV/$c$~\cite{radial}.

Investigations based on the TRENTO+MUSIC hydrodynamic framework have shown that bulk viscosity affects both the zero crossing and the high-$\pT$ behavior of $\vzpT$~\cite{v03,v02}. Including a temperature-dependent bulk viscosity improves agreement with the ATLAS data, although it shifts the zero crossing toward lower $\pT$. This shift signals a reduction of the mean $\pT$ in the presence of bulk viscosity. Before proceeding further, we note that in
macroscopic hydrodynamic models, the shear and bulk viscosity coefficients are explicit input parameters. AMPT, however, is a microscopic transport model in which the effective transport coefficients are emergent properties that arise dynamically from the microscopic interactions of the particles. Hence, we cannot tune the bulk viscosity directly as a standalone parameter in the present study. The effect of shear viscosity can be studied indirectly by changing the early-stage partonic scattering cross section ($\eta/s \propto 1/\sigma_{\rm parton}$) or the screening mass ($\mu$). Bulk viscosity, however, arises in kinetic theory from conformality breaking, for example through massive particles and inelastic scatterings. We therefore do not study the effect of bulk viscosity on $\vzpT$ at FAIR in the present work. Similarly, the effect of a suitable EoS, possibly that of baryon-rich QCD matter, can be studied in an appropriate hydrodynamic framework. We plan to investigate these issues in the future.



\begin{figure*}[t]
\centering
\includegraphics[width=\textwidth]{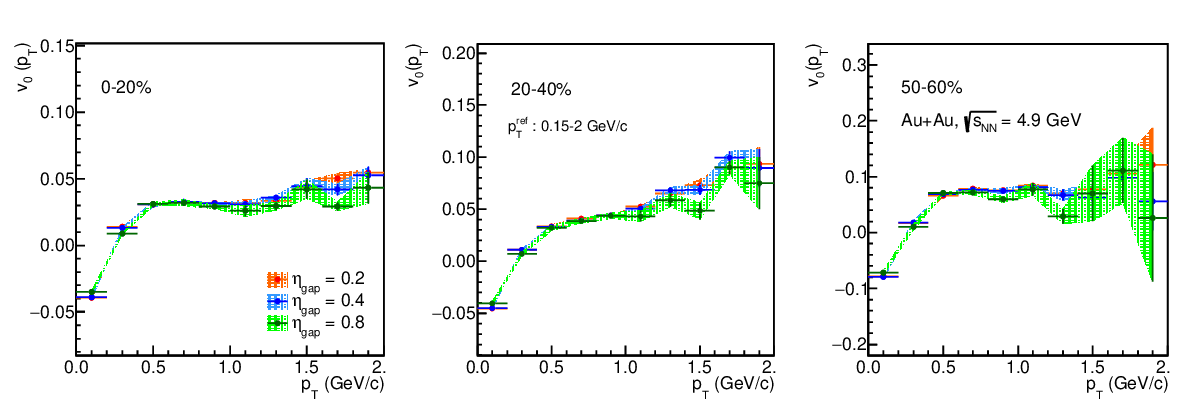}
\caption{(Color online) Pseudorapidity-gap dependence of $\vzpT$ for inclusive charged particles in Au+Au collisions at $\snn=4.9$~GeV for the 0--20\%, 20--40\%, and 50--60\% centrality classes, for the reference interval $0.15<\ptref<2.0$~GeV/$c$.}
\label{eps4}
\end{figure*}
The right panel of Fig.~\ref{eps2} presents $\vzpT\sqrt{\nch}$ as a function of $\pT$, where the charged-hadron $\vzpT$ in each centrality class is multiplied by $\sqrt{\nch}$ of that class. This representation tests whether the magnitude of the radial-flow fluctuations approximately follows an inverse-square-root multiplicity dependence, as predicted by hydrodynamic calculations~\cite{v01}. Although multiplying by $\sqrt{\nch}$ reduces the separation between the centrality classes in the low-$\pT$ region, the agreement deteriorates at higher $\pT$, suggesting that the scaling applies mostly to the bulk, low-$\pT$ region. This convergence at low $\pT$ indicates that a significant part of the centrality dependence of $\vzpT$ is related to multiplicity fluctuations varying as $1/\sqrt{\nch}$.

The ATLAS Collaboration at the LHC has previously studied $v_0$ as a function of the efficiency-corrected charged-hadron multiplicity $\nch$~\cite{v03}. 
For three selected $\ptref$ ranges, the corresponding $v_0$, obtained with $\Delta\eta=1$, shows scaling behavior as a function of $\nch$, and $\vzpT\sqrt{\nch}$ as a function of $\pT$ falls approximately on a common trend at low $\pT$.

\subsection{Centrality dependence of integrated radial-flow fluctuations}

Having discussed the $\pT$-differential radial-flow fluctuations, we now study the features of the $\pT$-integrated observable $v_0$. Fig.~\ref{eps3} (left) shows the centrality (upper left) and multiplicity (lower left) dependence of $v_0$ for different reference-$\pT$ intervals.
A clear, systematic increase of $v_0$ toward peripheral collisions is observed for all reference-$\pT$ intervals, although the absolute magnitude depends moderately on the selected $\ptref$ range.
These results are qualitatively consistent with the centrality dependence of $v_0$ measured by the ATLAS Collaboration in Pb+Pb collisions at $\snn = 5.02$~TeV~\cite{v03}.
They show that the relative event-by-event fluctuations of the mean transverse momentum become stronger as collisions become more peripheral. Since $v_0$ is defined as the relative RMS fluctuation of the event-wise mean $\pT$, an increase in $v_0$ directly indicates an increase in the relative fluctuations of the transverse-momentum distribution. The study at 200~GeV~\cite{radial2} similarly finds a monotonic increase of $v_0$ from central to peripheral collisions. The agreement between the reference intervals suggests that the centrality dependence is not an artifact of a particular $\ptref$ selection. In contrast to the centrality representation, $v_0$ decreases monotonically with increasing multiplicity, as expected.

The upper-right panel of Fig.~\ref{eps3} shows the normalized observable $v_0 / v_0^{0\text{--}5\%}$ as a function of centrality, where $v_0^{0\text{--}5\%}$ denotes the value in the most central collision class. The normalized curves overlap strongly for the different $\ptref$ selections.
Although the absolute $v_0$ depends on the selected momentum interval, its relative centrality evolution is almost independent of the reference-$\pT$ range, a feature also observed by the ATLAS Collaboration~\cite{v03}. This suggests that the different momentum selections respond to a common underlying event-by-event fluctuation of the overall spectral shape, i.e., the centrality evolution of radial-flow fluctuations is robust against the choice of reference momentum interval.



The four data sets in the lower-left panel of Fig.~\ref{eps3} are fitted with a power law, $v_0=A\,\nch^{b}$. Within the statistical precision of the present calculation, all four reference-$\pT$ intervals give an exponent consistent with $b\simeq-0.5$. The integrated radial-flow fluctuation therefore follows an approximate inverse-square-root multiplicity scaling, $v_0\propto \nch^{-1/2}$, very similar to what is observed at higher collision energies~\cite{v03,radial2}. This dependence can be understood qualitatively in terms of independent or weakly correlated particle-emitting sources: if the magnitude of event-by-event fluctuations decreases approximately as the inverse square root of the number of contributing sources, and the charged-particle multiplicity is a proxy for the effective number of sources, the observed scaling of $v_0$ is consistent with the dilution of relative mean-$\pT$ fluctuations as the number of produced particles increases. We note, however, that $v_0$ as defined in Eq.~(5) is computed within a single window and therefore also contains the purely statistical contribution of finite multiplicity, which by itself scales as $\nch^{-1/2}$; the observed exponent alone therefore does not separate dynamical from statistical fluctuations. The same multiplicity dependence is found for all four reference-$\pT$ selections. The results are consistent with those observed at higher energies.

The lower-right panel shows $v_0/v_0^{0\text{--}5\%}$ as a function of $\nch / \nch^{0\text{--}5\%}$.
Here the four reference-$\pT$ selections again collapse onto a common curve, reflecting their common $\nch$ dependence. Since each curve is normalized to its own 0--5\% value, this collapse follows largely from the common exponent and does not by itself probe the origin of the fluctuations.

\begin{figure*}[t]
\centering
\includegraphics[width=0.85\textwidth]{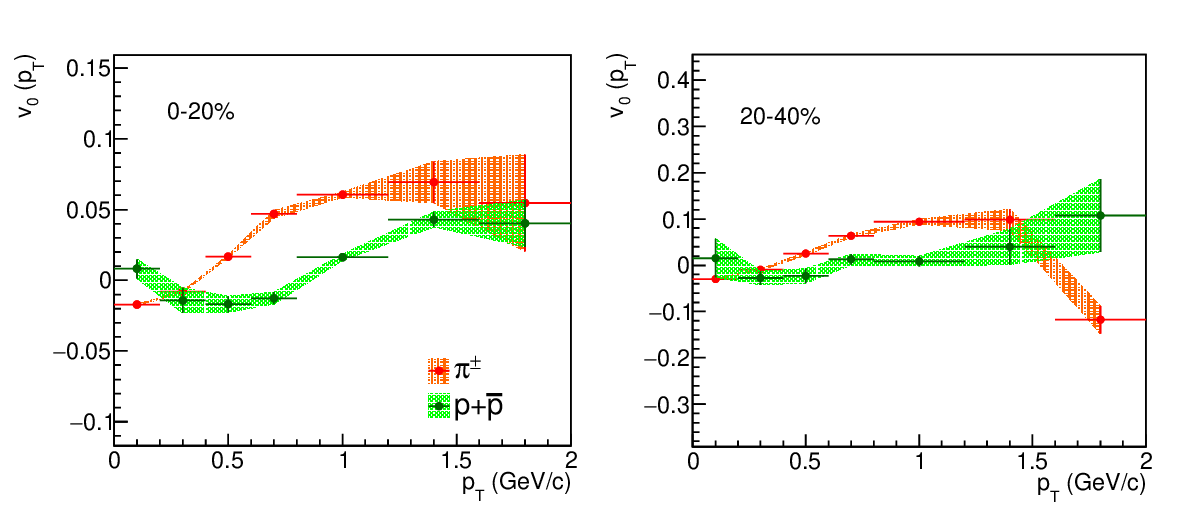}
\caption{(Color online) Transverse-momentum dependence of the radial-flow fluctuation observable $\vzpT$ for identified $\pi^\pm$ and $p+\bar p$ in Au+Au collisions at $\snn=4.9$~GeV for the 0--20\% and 20--40\% centrality classes.}
\label{eps5}
\end{figure*}

\begin{figure*}[t]
\centering
\includegraphics[width=0.85\textwidth]{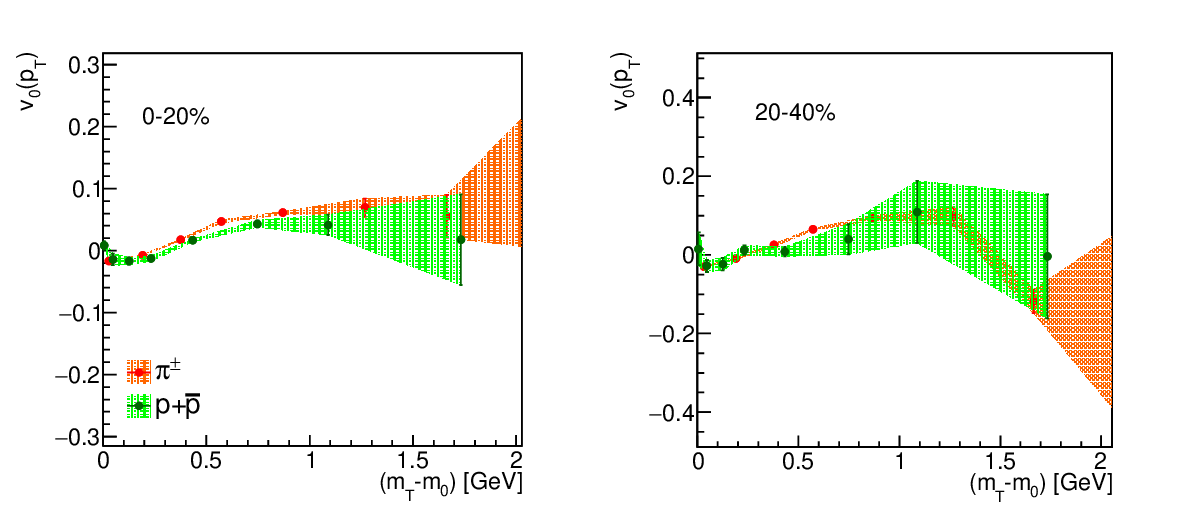}
\caption{(Color online) $\vzpT$ as a function of the transverse kinetic energy, $\mT-m_0$, for $\pi^\pm$ and $p+\bar p$ in Au+Au collisions at $\snn=4.9$~GeV for the 0--20\% and 20--40\% centrality classes.}
\label{eps6}
\end{figure*}

\subsection{$\eta$-gap dependence of $\vzpT$}
The influence of nonflow correlations on $\vzpT$ can be further tested by varying $\Delta \eta$ between the subevents. In Fig.~\ref{eps4}, we investigate the dependence of $\vzpT$ on the pseudorapidity separation between the two correlated regions for the three centrality classes and three values of the pseudorapidity gap, $\etagap=0.2$, 0.4, and 0.8. A larger $\Delta \eta$ suppresses short-range non-collective contributions originating from resonance decays, jet-like correlations, and other local particle-production mechanisms, but also amplifies flow-decorrelation effects~\cite{ATLAS:2020sgl}. For the gaps studied here, $\vzpT$ changes little with $\Delta \eta$ in all centrality intervals, which is compatible with the dominant contribution being a long-range correlation, as for anisotropic flow. Because the gap is varied only between 0.2 and 0.8 within $|\eta|\leq1$, this test cannot exclude nonflow correlations that extend over larger pseudorapidity ranges. The sensitivity of $\vzpT$ to the gap selection becomes pronounced only at larger $\pT$, above 1~GeV/$c$, suggesting a possible influence of short-range correlations or residual nonflow effects in the high-$\pT$ region.

Both LHC experiments have studied how $\vzpT$ of inclusive charged particles depends on the gap. ALICE varied $\Delta\eta$ from 0 to 1 in steps of 0.2 and found that, for $\pT < 3$~GeV/$c$, the result changes by less than 2\% in central and mid-central collisions and less than 8\% in peripheral collisions relative to $\Delta\eta = 0$, with differences of up to about 15\% at higher $\pT$ that indicate a modest nonflow contribution~\cite{radial}. ATLAS, varying the gap from 0 to 3, saw little change in central collisions and a small reduction in peripheral collisions above 3~GeV/$c$, which it attributes to residual nonflow from away-side jets~\cite{v03}.

\begin{figure*}[t]
\centering
\includegraphics[width=0.85\textwidth]{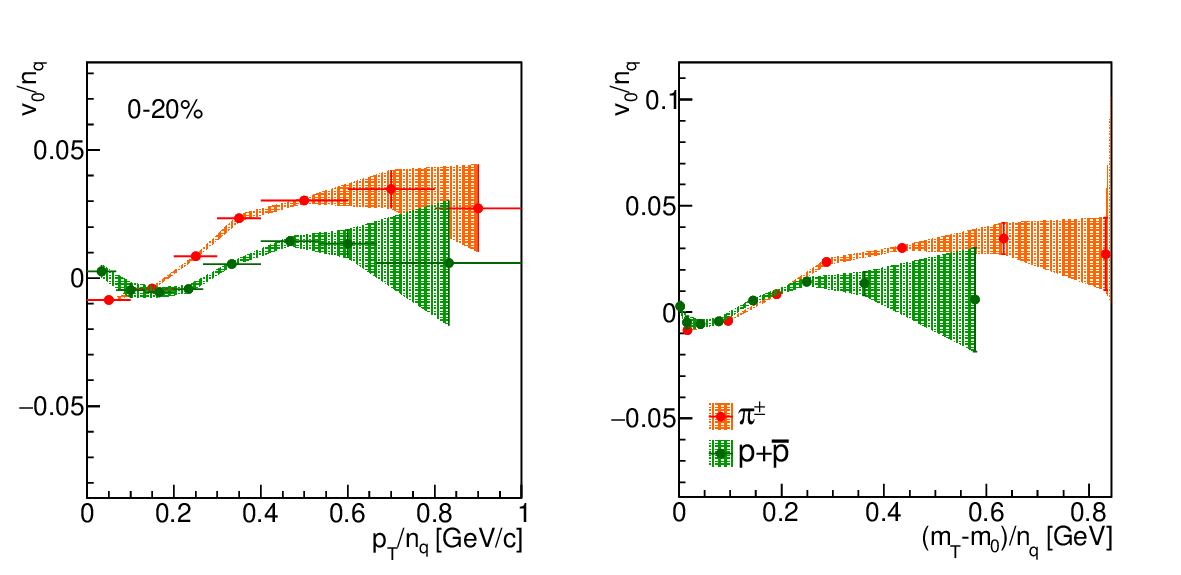}
\caption{(Color online) NCQ-scaled radial-flow fluctuation observable, $\vzpT/n_q$, as a function of $\pT/n_q$ (left) and $(\mT-m_0)/n_q$ (right) for $\pi^\pm$ and $p+\bar p$ in 0--20\% central Au+Au collisions at $\snn=4.9$~GeV.}
\label{eps7}
\end{figure*}
\subsection{$\vzpT$ for identified charged hadrons}

Having studied $\vzpT$ for inclusive charged particles, we now turn to identified hadrons. Fig.~\ref{eps5} shows $\vzpT$ as a function of $\pT$ for $\pi^\pm$ and $p+\bar p$ produced in Au+Au collisions at $\snn = 4.9$~GeV, for the 0--20\% and 20--40\% centrality intervals. The spectator protons are excluded from our study by applying a cut on $p_{T}$. While the inclusive $\vzpT$ characterizes the overall fluctuation of the $\pT$ spectrum, the identified-particle results show how this fluctuation is mapped onto particles of different masses. In the present calculation, the pion and proton $\vzpT$ exhibit a pronounced species dependence. Above $\pT\simeq 0.4$~GeV/$c$, the pion $\vzpT$ is larger than the proton value over most of the calculated range; in the 20--40\% class the highest-$\pT$ pion point falls below the proton point. This ordering, $v_0^{\pi}(\pT)>v_0^{p}(\pT)$ at fixed $\pT$, is qualitatively similar to the hydrodynamic mass ordering of radial-flow fluctuations seen at higher energies. However, the baryon--meson crossing at intermediate $\pT$, seen at higher collision energies and expected from the collective radial expansion of the produced medium in a hydrodynamic framework~\cite{radial,hydro,radial2}, is absent here.

Fig.~\ref{eps6} presents $\vzpT$ as a function of transverse kinetic energy, $\mT-m_0$, where $\mT=\sqrt{\pT^2+m_0^2}$ denotes the transverse mass and $m_0$ is the rest mass of the corresponding hadron species. Expressing the observable in terms of $\mT-m_0$ allows a more direct comparison of the radial dynamics at a common transverse kinetic-energy scale. At low transverse kinetic energy, the pion and proton distributions are considerably closer than when plotted as a function of $\pT$; the strong separation between the two species in the $\pT$ representation is reduced. This indicates that part of the apparent species dependence of $\vzpT$ arises from the different particle masses and the corresponding mapping between $\pT$ and transverse kinetic energy.

Number-of-constituent-quark (NCQ) scaling is regarded as one of the key signatures of the formation of a strongly interacting partonic medium~\cite{scaling1,scaling2,scaling4}. The mass ordering of $\vzpT$, together with the approximate convergence of the pion and proton distributions as a function of $\mT-m_0$, is consistent with collective radial dynamics. To investigate whether the observed behavior exhibits the constituent-quark scaling associated with partonic collectivity, the observable is further examined after scaling by the number of constituent quarks, $n_q$, as shown in Fig.~\ref{eps7}. In contrast to the results 
obtained at higher collision energies, the pion and proton distributions in Au+Au collisions at $\snn=4.9$~GeV do not collapse onto a common curve after $n_q$ scaling; a clear separation between the two species remains.

At the LHC, ALICE measured $\vzpT$ of pions, kaons, and protons up to $\pT = 6$~GeV/$c$~\cite{radial}. The three species are mass ordered below 3~GeV/$c$ in every centrality class, while above 3~GeV/$c$ the proton values exceed those of the mesons. After NCQ scaling the species approximately coincide, which ALICE takes as evidence that collectivity develops among partons before hadronization and that hadrons form by coalescence of flowing quarks. AMPT-SM calculations at $\snn = 200$~GeV show the same baryon--meson splitting and NCQ scaling~\cite{radial2}.

The absence of NCQ scaling in the AMPT-SM results in the FAIR energy domain indicates that the calculated radial-flow fluctuations cannot be interpreted as evidence for a universal constituent-quark-level collective response. In particular, the present results do not support the interpretation that the radial-flow fluctuations are dominated by partonic collectivity, as observed at higher energies~\cite{radial,radial2}. In heavy-ion collisions, with decreasing beam energy, the lifetime of the partonic phase becomes shorter~\cite{Sahoo:2010gr}. Thus, one possible explanation for the violation of NCQ scaling, which we have not explicitly tested here, is that the partonic stage in low-energy collisions at FAIR is too short-lived for collective motion to develop among partons; hadronic rescattering after coalescence could also wash out a partonic scaling. 
In addition, at low collision energies, the high-$\mu_{B}$ matter produced by baryon stopping, which transports entrance-channel quarks to midrapidity, can also result in the breaking of NCQ scaling~\cite{Dunlop:2011cf}.
Calculations with the hadronic cascade switched off, or with a larger parton scattering cross section, would distinguish some of these possibilities.

\begin{figure*}[t]
\centering
\includegraphics[width=1.01\textwidth]{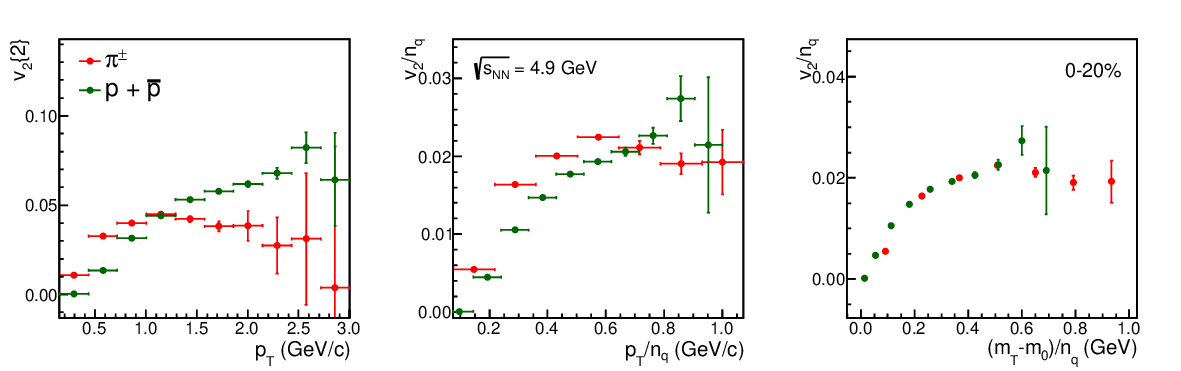}
\caption{(Color online) $p_T$-dependent elliptic flow $v_2$ (left), $v_2/n_q$, as a function of $\pT/n_q$ (middle) and $(\mT-m_0)/n_q$ (right) for $\pi^\pm$ and $p+\bar p$ in 0--20\% central Au+Au collisions at $\snn=4.9$~GeV.}
\label{eps8}
\end{figure*}

Before we close, it may be noted that in Ref.~\cite{Zhu:2025kud}, the authors have employed the AMPT model to study the NCQ scaling of $v_2$ of identified hadrons in non-central Au+Au collisions at $\sqrt{s_{NN}} = 3.0$--7.7~GeV. Interestingly, in the SM mode, NCQ scaling is seen to be present for $\sqrt{s_{NN}} \ge 3.9$ GeV and violated at $\sqrt{s_{NN}} = 3$ GeV. However, the scaling behavior can be gradually recovered by artificially increasing the parton scattering cross section from 3 mb to 200 mb. In order to test the consistency of our results, we have explicitly investigated the $v_{2}(p_{T})$ and NCQ scaling behavior for 0--20\% central collisions. Results are depicted in Fig.~\ref{eps8}. As evident, a clear crossover with proton $v_{2}(p_{T})$ surpassing the pion $v_{2}(p_{T})$ is observed around $p_{T} \simeq 1.2$~GeV/$c$. The $v_{2}/n_{q}$ exhibits reasonable scaling behavior as a function of $(\mT-m_0)/n_q$. Further studies are required to understand the reason for the presence of NCQ scaling in $v_{2}(p_{T})$ and its absence in $v_{0}(p_{T})$ of identified hadrons within the AMPT-SM model at $\sqrt{s_{NN}} = 4.9$ GeV. This also highlights the need to measure both the observables, $v_{0}(p_{T})$ and $v_{2}(p_{T})$, to investigate the collectivity at FAIR. In addition, the rapidity dependence of radial-flow fluctuations, $v_{0}(\eta)$, has recently been proposed as a promising observable to probe the EoS of baryon-rich matter, having negligible influence from shear and bulk viscosities as well as baryon diffusion. A rapidity-dependent splitting of $v_{0}$ between protons and anti-protons indicating different radial dynamics of baryons and anti-baryons in a dense baryonic medium has been predicted by hybrid model simulations in $\snn=19.6$ GeV Au+Au collisions~\cite{Parida:2026kyo}. It will also be interesting to explore such rapidity-dependent $v_0$ splitting in the FAIR energy domain. 

\section{Summary and Conclusions}
We have studied $\vzpT$, the $\pT$-differential measure of radial-flow fluctuations, in Au+Au collisions at $\snn = 4.9$~GeV with the AMPT-SM model. Unlike blast-wave fits, which give a single $\pT$-integrated flow velocity from inclusive spectra, $\vzpT$ resolves the $\pT$ dependence of radial flow and, through a pseudorapidity gap, suppresses short-range nonflow. At the LHC, ALICE and ATLAS measurements in Pb+Pb collisions at $\snn = 5.02$~TeV have established the main collective features of $\vzpT$---gap independence, factorization, a centrality-independent shape, mass ordering, and NCQ scaling---as well as its sensitivity to bulk viscosity, the EoS, and initial-state fluctuations~\cite{v03,radial}. Published measurements so far exist only at that energy; preliminary STAR results extend them to Au+Au collisions at $\snn = 7.7$--200~GeV~\cite{STARprelim}.

No other study of $\vzpT$ is available at FAIR energies. The calculated $\vzpT$ exhibits trends similar to those observed at higher energies. For all centrality classes investigated, $\vzpT$ changes from negative at very low $\pT$ to positive at higher $\pT$. It fluctuates more at high $\pT$ and in peripheral collisions, where the available multiplicity is smaller and the statistical uncertainties are substantially larger. The dependence of $\vzpT$ on the choice of reference-$\pT$ interval provides an additional test of the robustness of the observable. The extracted $\vzpT$ changes little when the edges of the reference-$\pT$ interval are varied, in the bulk region; a stronger test would use disjoint reference intervals.

The integrated $v_0$ shows a pronounced centrality dependence, increasing continuously from central to peripheral collisions, accompanied by a corresponding decrease of $v_0$ with increasing charged-particle multiplicity $\nch$. The results for the different reference-$\pT$ selections follow a nearly common trend. The observed dependence is approximately consistent with an inverse-square-root multiplicity scaling. This indicates that the dominant centrality dependence of the integrated fluctuation signal is closely related to the number of particles participating in the correlation.

The results also change little for $\etagap$ between 0.2 and 0.8, so the calculated signal is not dominated by short-range correlations within that range. The pion and proton $\vzpT$ show an ordering at low $\pT$ qualitatively similar to the mass ordering seen at higher energies; however, the absence of constituent-quark scaling suggests that the collective behavior is unlikely to originate from partonic collectivity, as observed at higher collision energies.

Overall, the persistence of a common trend, the approximate multiplicity scaling, the stability with respect to the reference-$\pT$ interval, and the weak dependence on the pseudorapidity gap, all similar to higher energies, are consistent with a long-range, bulk-dominated component of the correlations in AMPT-SM at $\snn=4.9$~GeV; however, they do not justify claiming the development of collectivity at the partonic level. These results can serve as a useful reference for future measurements once the CBM experiment becomes operational. Establishing a firm connection between these observations and collective radial dynamics will require comparisons with a non-collective baseline and with complementary observables, as well as future measurements at CBM. Such studies are left for future investigation.

\begin{acknowledgments}
We acknowledge the VECC grid computing facility for the event generation. We thank Dr.~Vikas Singhal, Mr.~Prasun Singha Roy, and Mr.~Monirul Purkait for their help and support in data generation by providing the required computing facilities. We also thank Dr.~Tribhuban Parida for many stimulating discussions and Prof.~C.~M.~Ko and Prof.~Z.~W.~Lin for useful discussions regarding AMPT.
\end{acknowledgments}

\end{document}